\documentclass[12pt]{article}
\usepackage{lineno}

\usepackage[margin=1in]{geometry}
\usepackage{amssymb,amsmath,latexsym}                        
\usepackage{graphicx}
\usepackage{cite}
\usepackage{xcolor}
\usepackage{float}

\usepackage{amsthm}
\usepackage{multicol}
\usepackage{authblk}
\usepackage[colorlinks]{hyperref}

\usepackage{adjustbox}
\usepackage{bbm}
\usepackage{mathtools}

\date{} 
\usepackage{graphicx}
\usepackage{subcaption}
\begin{document}

\title{\boldmath A standalone simulation framework of the MRPC detector read out in waveforms}
\author[a]{Fuyue Wang}
\author[a]{Dong Han}
\author[a]{Yi Wang\footnote{Corresponding author. Email: yiwang@mail.tsinghua.edu.cn.}}
\author[a]{Yancheng Yu}
\author[a]{Qiunan Zhang}
\author[b]{Baohong Guo}
\author[a]{Yuanjing Li}

\affil[a]{\normalsize\it Tsinghua University, \\Key Laboratory of Particle and Radiation Imaging, Ministry of Education, Beijing 100084, China}
\affil[b]{\normalsize\it Nuctech Company Limited}

\maketitle

\begin{abstract}

Dedicated detector simulations are very important for the success of high energy physics experiments. They not only bring benefits to the optimization of the detectors, but can also improve the precision of the physics results. To design a Multi-gap Resistive Plate Chamber (MRPC) with a very good time resolution, a detailed monte-carlo simulation of the detector is needed and described in this paper. The simulation can produce detector signal waveforms which contain a complete information about the events. The detector performance filled with different gas mixtures is studied, and comparison between simulation and experimental results show a good agreement. 
\end{abstract}

%\keywords{MRPC, simulation framework, waveform, time resolution}

%\proceeding{N$^{\text{th}}$ Workshop on X\\
%  when\\
%  where}
  
% Simulation is very important either in standalone detector study and 
%-------------------------------------------------------------------------------
\section{Introduction}
\label{sec:intro}
Accurate detector modeling and detailed monte-carlo simulations have played an important role in the success of the high energy physics experiments through the past years. At the Large Hadron Collider(LHC) at the European Center of Nuclear Research(CERN), the simulation benefits not only the design and optimization of the detectors, the development of the software\cite{elvira2017impact}, but also the data processing and the track reconstruction algorithms\cite{Atlasneural} which further improve the precision of the physics results.

In early times, detector simulation was only based on a few simple analytical equations and used as a qualitative proof of large physics effects or biases. Detailed monte-carlo(MC) detector simulations started from about 1980s, when the Electron Gamma Shower software(EGS)\cite{egs2006} and GEANT3\cite{geant1987} were released by their authors. The MC simulation accurately models the geometry of the detector, propagates the particles through the detector and simulates the interactions between particles and the surrounding material. The goal of the simulation is to produce the same events in the scenario of the actual experiment. Hence the output of the simulation usually takes the same forms as the data collected from the experiment. 

In this paper, we focus on the simulation of the Multi-gap Resistive Plate Chamber(MRPC), which was proposed in 1990s and originated from the Resistive Plate Chamber(RPC). The simulation of the RPC was studied and improved by many previous authors\cite{zeballos1996avalanche,abbrescia1997properties,abbrescia1999progresses,riegler500144} from the 1990s, and based on this, \cite{an2016monte} has proposed a simulation code for MRPC in the framework of the BESIII experiment. However a dedicated standalone simulation of the MRPC is very rare, and therefore it is proposed in this paper. Most present MRPCs used in high energy experiments are equipped with a front-end electronics that only provides the information of $t_c$(the time when the signal overcomes a fixed threshold) and $t_{tot}$(the time interval the signal stays above that threshold, usually called time over threshold). With progresses in physics and the increase of the beam energy, a higher demand for the time resolution is raised. In these conditions, the information of the entire signal waveform is valuable because it provides more information than just the $t_c$ and $t_{tot}$, so it is included in the simulation.

A typical simulation of the high energy physics detectors consists of the following modules: 
\begin{enumerate}
\item is the simulation of the particle source which contains the type, the position and the angle of the incident particles. 
\item is the simulation of the particles impinging on the detector and the primary interactions between the particles and the matter. The first and second modules are generally based on the Geant4 package\cite{Agostinelli:2002hh}.
\item The primary energy deposited in the detector ionizes the molecules of the detector material and creates electrons and positive ions(or holes). The drift and, if the electric field is intense enough, the avalanching of the electrons induce a current signal on the readout electrodes. This module contains the simulation of the drifting and the signal induction. 
\item simulates the electronics response to the induced current signal. 
\end{enumerate}
The first 2 modules are described in Sec.\ref{sec:primary} and the last 2 modules are in Sec.\ref{avalanche}. The results of the simulation and the comparison with the experiment are shown in Sec.\ref{sec:ressim}.

\section{Detector geometry and primary energy deposition}
\label{sec:primary}
\begin{figure*}[h!]
	\centering
	\includegraphics[width=0.5\textwidth]{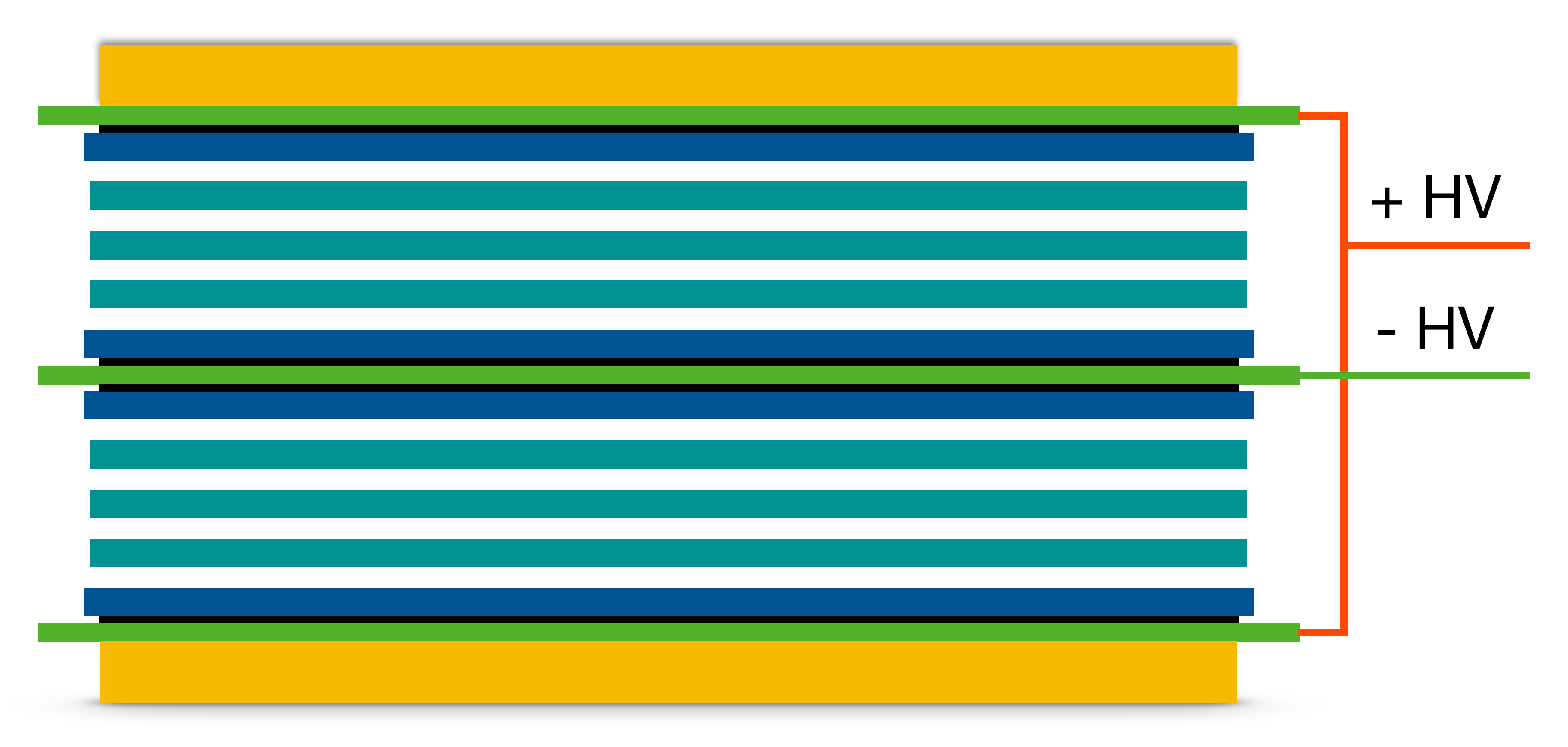}
	\caption{The structure of the detector}
	\label{fig:geo}
\end{figure*}

In this simulation code, the source of the particles can be defined to be of any type, placed at any position and shot at any direction. To compare with the cosmic ray test conducted in our laboratory, we simulated muons with an energy of 4 GeV\cite{pdg4GeVmuon} and incidence angle perpendicular to the detector with an uniform smear of 1 mrad. The detector geometry in the simulation is shown in Fig.\ref{fig:geo}. It has 2 stacks and each stack has 4 0.25-mm-thick gaps. The resistive plates are 0.7 mm thick and they are made of low resistive glass with a bulk resistivity of $10^{10}$ $\rm \Omega\, cm$. More details about the glass can be found in \cite{wang2010development,CBMTDR}. The first and the third electrodes are connected to the positive high voltage, while the second is negative. The electrodes are also made of 0.7mm-thick low resistive glass but coated with thin graphite layers. To test the accuracy of the simulation, we simulate 2 sets of gas mixtures: 1) 90\% $\rm C_2H_2F_4$, 5\% $\rm C_4H_{10}$ and 5\% $\rm SF_6$, 2) 95\% $\rm C_2H_2F_4$ and 5\% $\rm C_4H_{10}$. The working gas is at room temperature and under standard atmosphere.

\begin{figure*}[h!]
	\centering
	\includegraphics[width=0.5\textwidth]{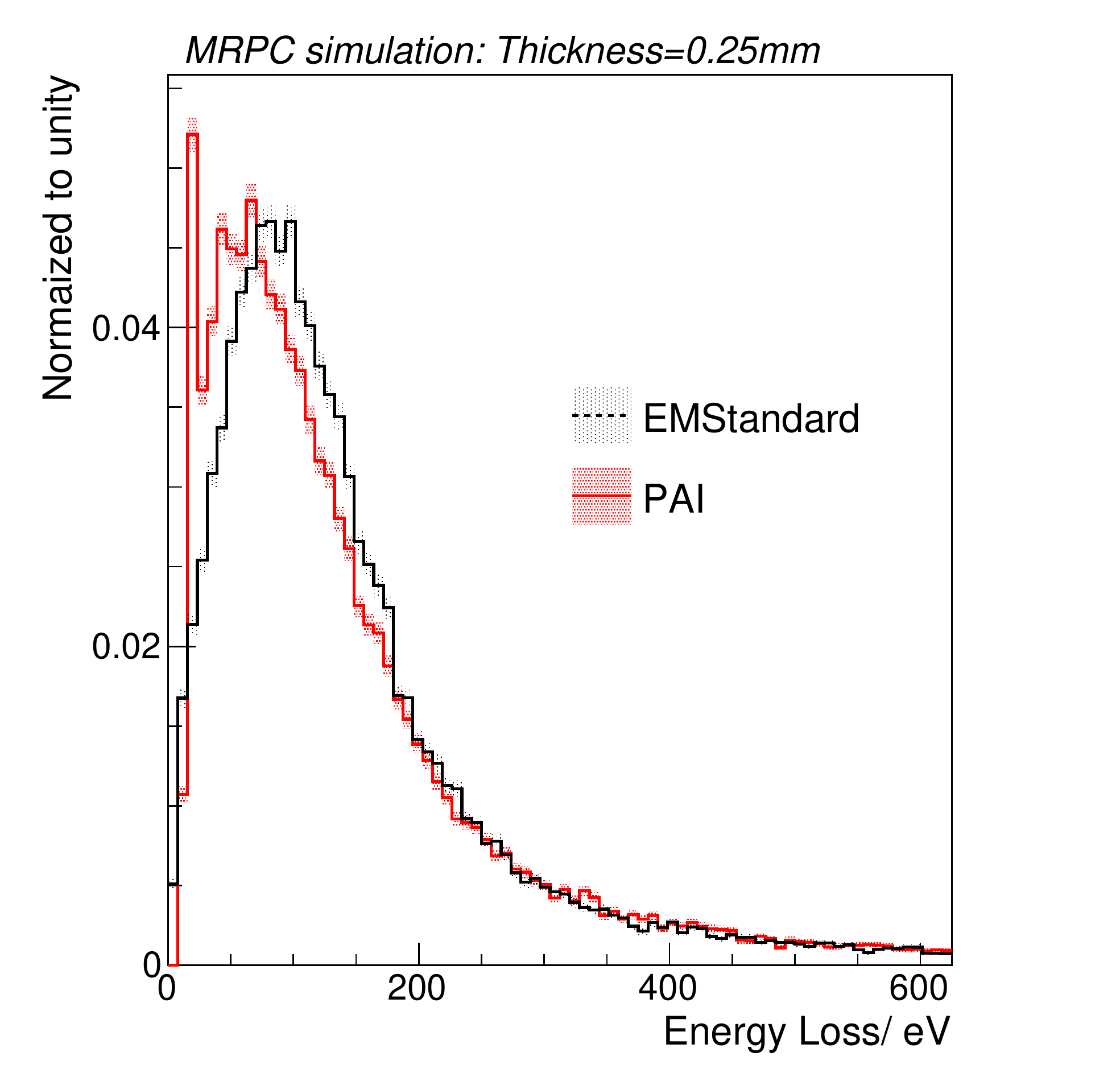}
	\caption{The distribution of the energy deposition in a 0.25 mm gap given by \texttt{EMstandard} and the \texttt{PAI} model.}
	\label{fig:model}
\end{figure*}

When the muons pass through the detector, they interact with the material and leave energy inside the detector. In the LHC simulation framework, the energy deposition is usually simulated based on the \texttt{EMstandard} physics process list implemented in Geant4 package\cite{ALLISON2016186}. However, it does not include the shell electron effect and is only excellent for thick sensors\cite{Wang20181}. Geant4 also includes a more detailed energy loss model: the Photo Absorption Ionization (\texttt{PAI}) model\cite{Apostolakis:2000yu}. This model is based on a corrected table of the photo-absorption cross section coefficients and works for various elements. The energy loss provided by this model is proved to be in a good agreement with the experiment data for thin sensors\cite{Allison:1980vw}. Fig.\ref{fig:model} shows the distribution of the energy deposition in a 0.25 mm thick gap given by both the \texttt{EMstandard} and the \texttt{PAI} model. Differences exist in the low energy region due to the binding energy of shell electrons of the detector material. Since the \texttt{PAI} model is more accurate for thin sensors, it is chosen in this simulation, but the \texttt{EMstandard} is also provided in this framework.
%The 3 peaks in the low energy region of the \texttt{PAI} model correspond to the binding energy of K, M and L shell electrons of the detector material. 

\section{Avalanche Multiplication and Signal formation}
\label{avalanche}
\begin{figure*}[h!]
	\centering
	\includegraphics[width=0.5\textwidth]{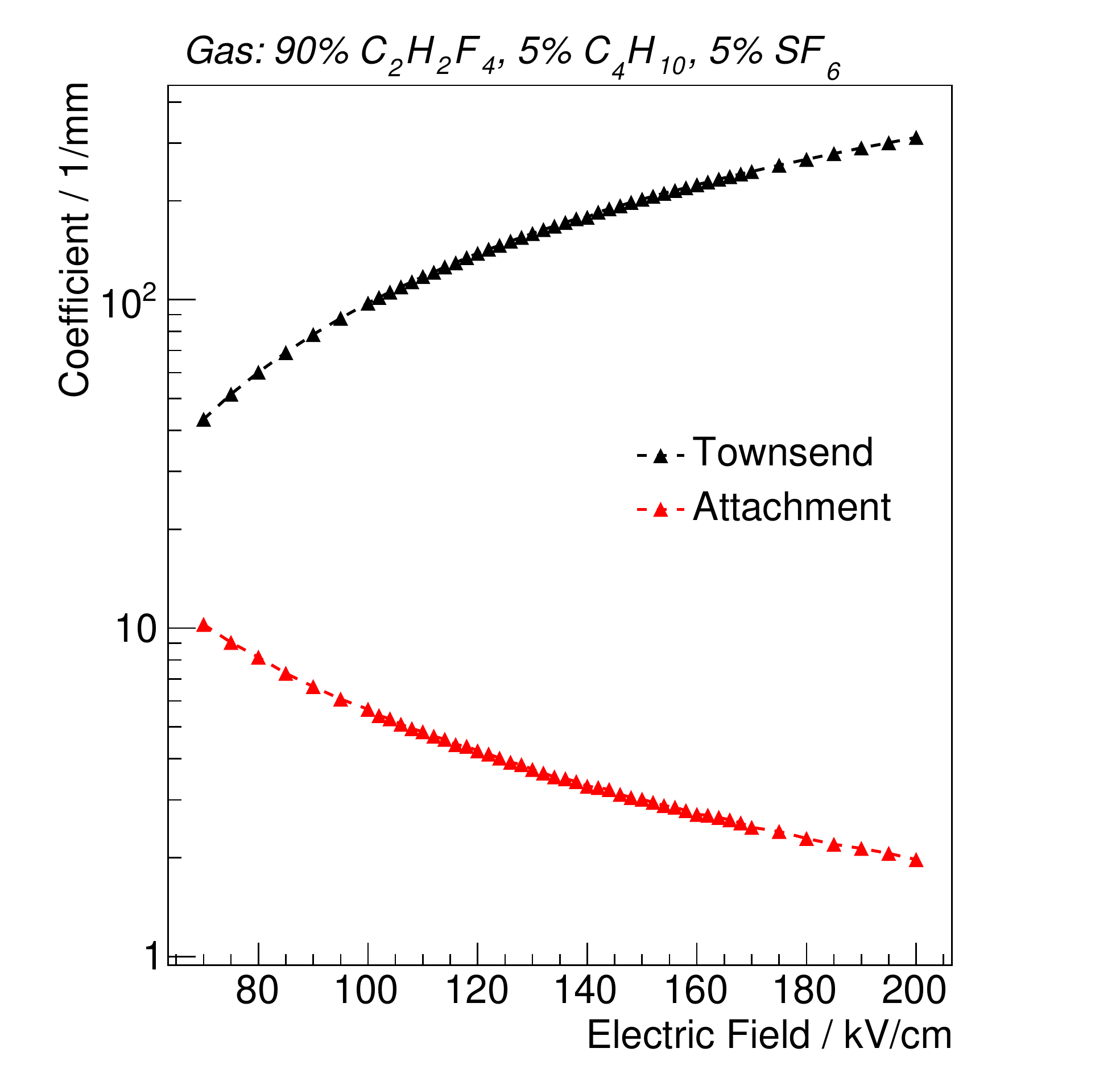}
	\caption{Townsend coefficient and attachment coefficient calculated by Magboltz\cite{magboltz1997}.}
	\label{fig:GasPara}
\end{figure*}

in the simulation, the energy deposited by the primary interactions in the gas gap ionizes electron-ion pairs with an average energy of 20 eV per pair\cite{an2016monte}. The electrons, once created, drift to the anode and start the avalanche multiplication under the applied electric field. We assume that the electric field in the gap is uniform and the electrons multiplies independently of the position and multiplication in the last step. The avalanche development follows Townsend equation\cite{townsend1900} reported here:
\begin{align}
\label{eq:townsend}
	\frac{d\bar{n}}{dx}=(\alpha-\eta)\bar{n},
\end{align}	
\noindent where $\bar{n}$ is the number of electrons at position $x$, $\alpha$ is the Townsend coefficient, $\eta$ is the attachment coefficient, and $\alpha-\eta$ is the first effective Townsend coefficient. Fig.\ref{fig:GasPara} shows the value of these two parameters with respect to the electric field computed using the Magboltz\cite{magboltz1997} software. Considering all the possible cases, the probability $P(n,x+dx)$ for an avalanche started with a single electron at position 0 to contain $n$ electrons after a distance $x+dx$ should be \cite{riegler500144}:
\begin{equation}
\label{eq:prob}
\begin{split}
P(n,x+dx)=&P(n-1,x)(n-1)\alpha \,{\rm d}x(1-(n-1))\eta \,{\rm d}x\\
&+P(n,x)(1-n\alpha \,{\rm d}x)(1-n\eta \,{\rm d}x)\\
&+P(n,x)n\alpha \,{\rm d}x\,n\eta \,{\rm d}x\\
&+P(n+1,x)[1-(n+1)\alpha \,{\rm d}x](n+1)\eta \,{\rm d}x
\end{split}
\end{equation}
\noindent The first line is the probability when there are $n-1$ electrons at $x$, one of them duplicates and none of them is attached. The second line is when there are $n$ electrons at $x$, none of them duplicates and none of them is attached. The third line is when there are also $n$ electrons at $x$, one of them duplicates and one of them is attached. The last line is the case when there are $n+1$ electrons at $x$, none of them duplicates but one is attached. A complete description of the solution to Eq.\ref{eq:prob} can be found in \cite{riegler500144}. We divide every gas gap into $N$ steps of $\Delta x$, and simulate the multiplication in every single step starting from the position of the primary interactions. It is proved that, after some steps the avalanche grows smoothly like $e^{(\alpha-\eta)x}$. When the number of the electrons is sufficiently large, the electric field produced by the avalanche cluster is comparable to the applied field, and thus the real electric field $E$ seen by the electrons decreases and this is called the space charge effect. Assuming that the cluster of electrons grows like a circle in the transverse direction, and the transverse diffusion length is around 100 $\rm\mu m/\sqrt{cm}$ at $E\approx100$ kV/cm\cite{lippmannthesis2003}, a cluster of $10^6$ electrons produces an electric field around 50 kV/cm at its surface. Therefore in the simulation, we limit the size of the cluster electrons to be around $10^6$ like it was done in \cite{abbrescia1999progresses}. The specific value varies a little with respect to the electric field\cite{spacecharge1998streamer}.
\begin{figure*}[h!]
	\centering
	\includegraphics[width=0.45\textwidth]{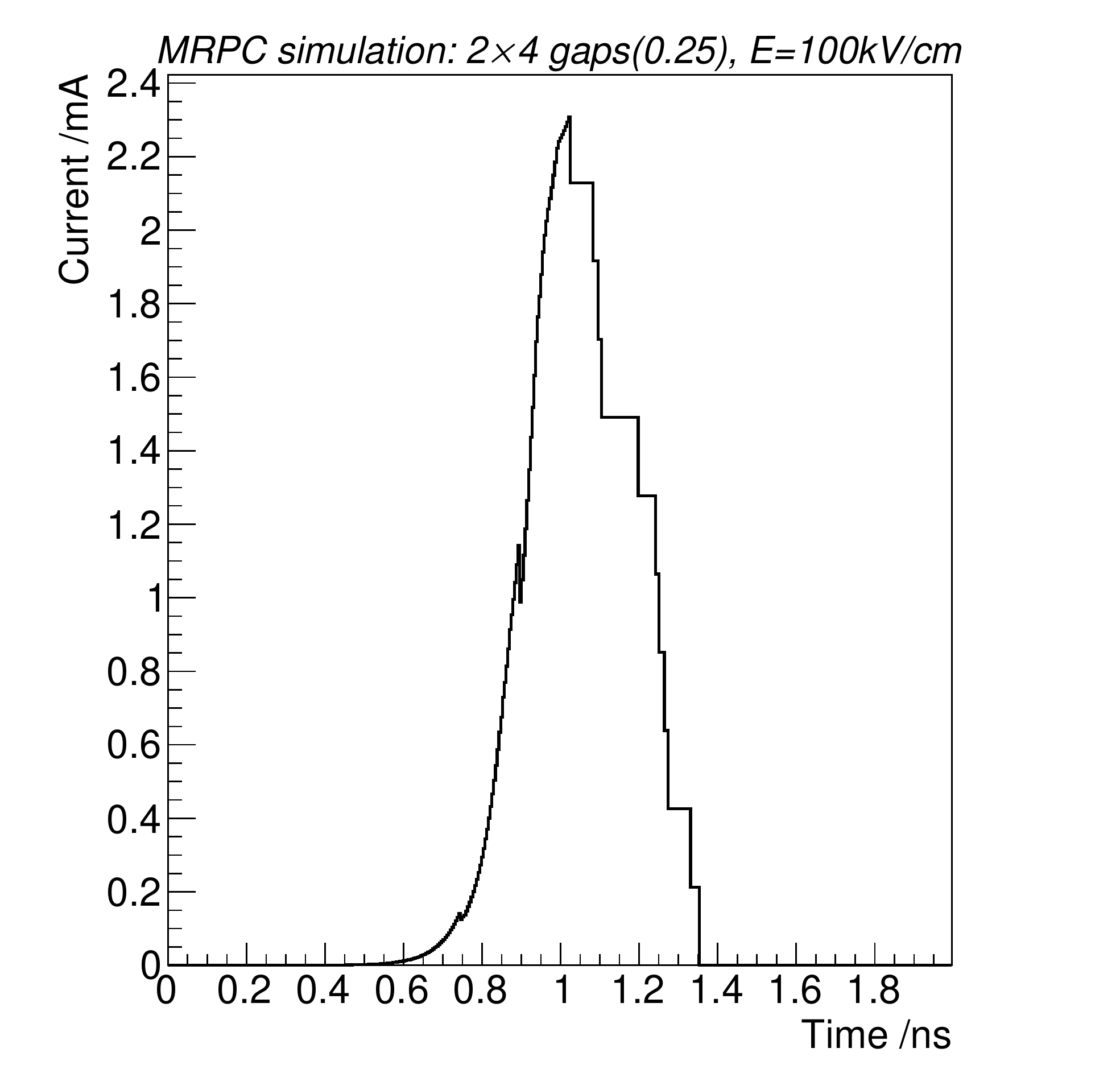}
	\hspace{0.1in}
	\includegraphics[width=0.45\textwidth]{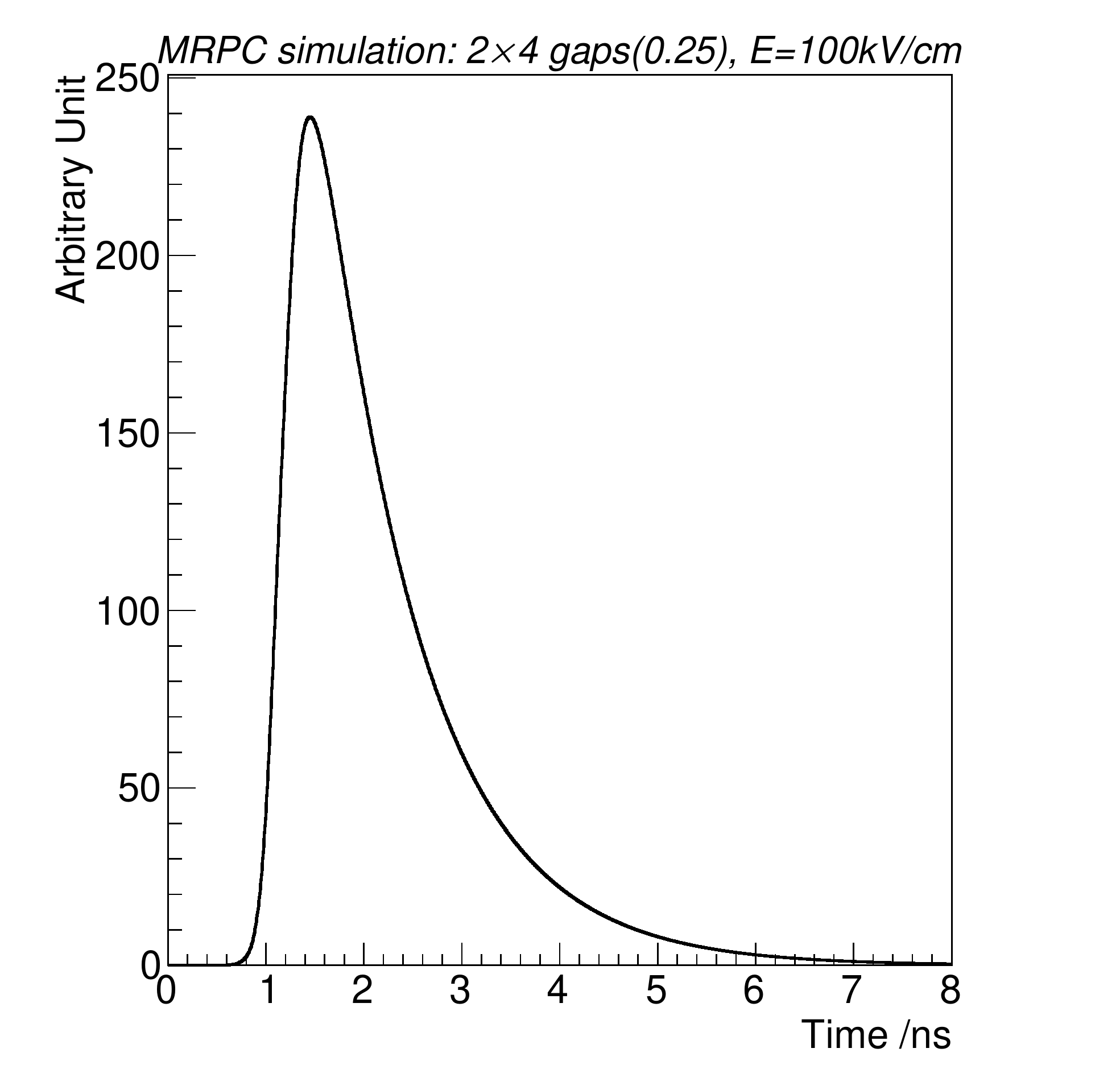}
	\caption{The signal of MRPC. (a)The signal induced on readout electrodes; (b)The signal convoluted with readout electronics response.}
	\label{fig:signal}
\end{figure*}

The drift of the multiplied electrons in the electric field induces a current signal on the read out electrodes. For the MRPC detector, we neglect the signal induced by the drift of the ions because the velocity of ions is slower than electrons by three orders of magnitudes. According to Ramo theory\cite{ramo1939}, the induced current is:
\begin{equation}
\label{eq:induce}
i(t)=\frac{\mathbf{E_W}\cdot \mathbf{v}}{V_W}e_0N(t)
\end{equation}
\noindent where $\mathbf{v}$ is the drift velocity, $e_0$ is the electron charge, $N(t)$ is the number of electrons at time $t$, $\mathbf{E_W}$ is the weighting field which is the electric field when setting the potential of the read out electrode to be $V_W$ and others 0. For the MRPC studied in this paper, the value of the weighting field under the center of the readout strip is $E_W/V_W=0.71$ mm$^{-1}$, according to the equation given in \cite{riegler500144}. Fig.\ref{fig:signal}(a) shows the signal induced on readout electrodes. We include the front-end electronics(FEE) response to the induced current by convolving Eq.\ref{eq:induce} with a simplified electronics response $f(t)$:
\begin{equation}
\label{eq:FEE}
f(t)=A(e^{-t/\tau_1}-e^{-t/\tau_2})
\end{equation}
\noindent where $A$ is the amplification factor, and $\tau_1$, $\tau_2$ correspond to the time constants of the RC circuits in the electronics. The length of the leading and trailing edge has a positive correlation with the value of $\tau_1$ and $\tau_2$, and thus these values should be calibrated with the experiments. Fig.\ref{fig:signal}(b) shows an example of the read out signal by this FEE without the electronics noise. The noise is introduced by adding a Gaussian(0,$\sigma$) random number to the signal in every time bin. $\sigma$ is the equivalent noise charge at the output. 
\begin{figure}
    \centering
    \begin{subfigure}[b]{0.45\textwidth}
        \includegraphics[width=\textwidth]{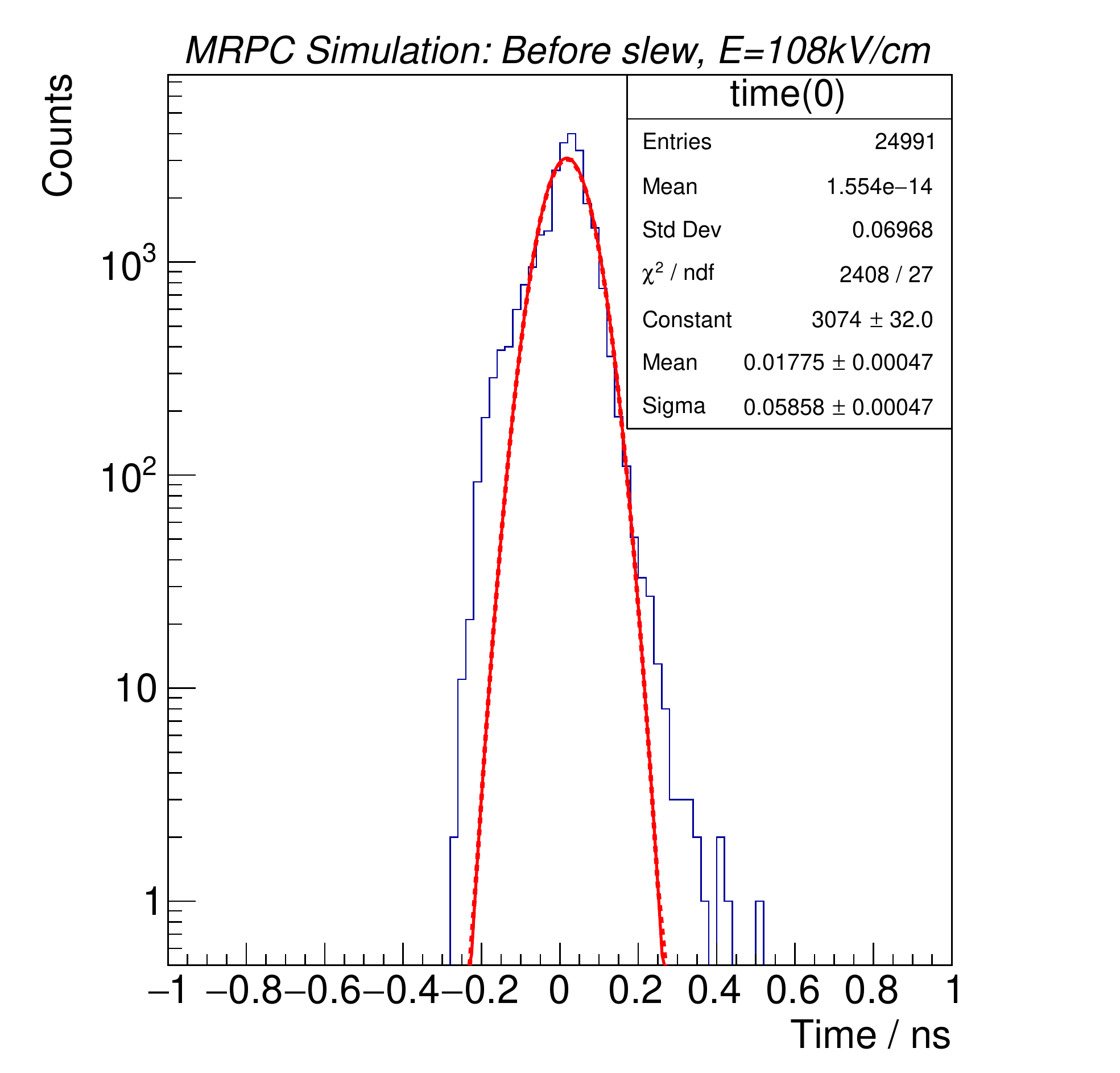}
        \caption{}
        \label{fig:slewa}
    \end{subfigure}
    \begin{subfigure}[b]{0.45\textwidth}
        \includegraphics[width=\textwidth]{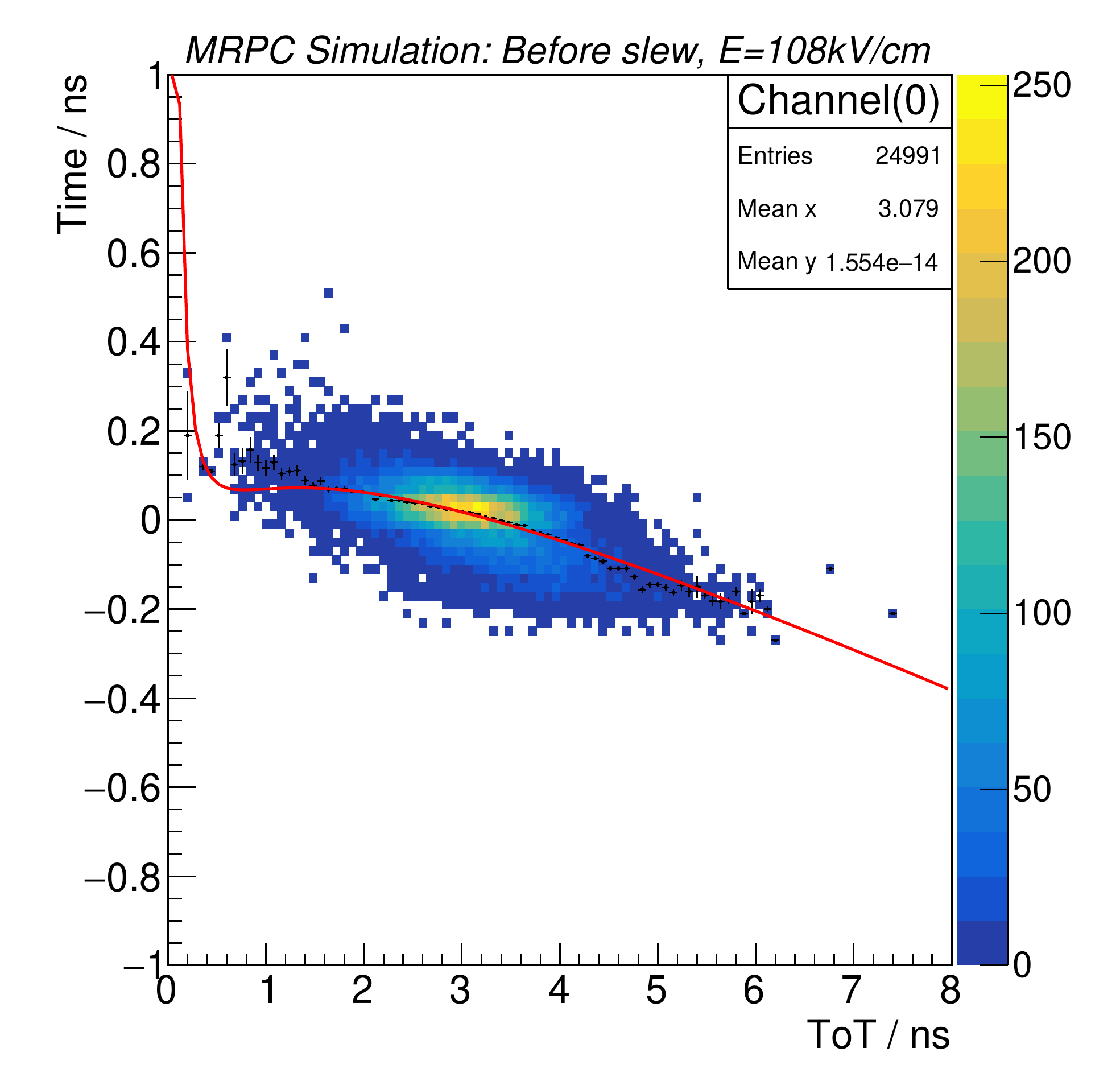}
        \caption{}
        \label{fig:slewb}
    \end{subfigure}
    \begin{subfigure}[b]{0.45\textwidth}
        \includegraphics[width=\textwidth]{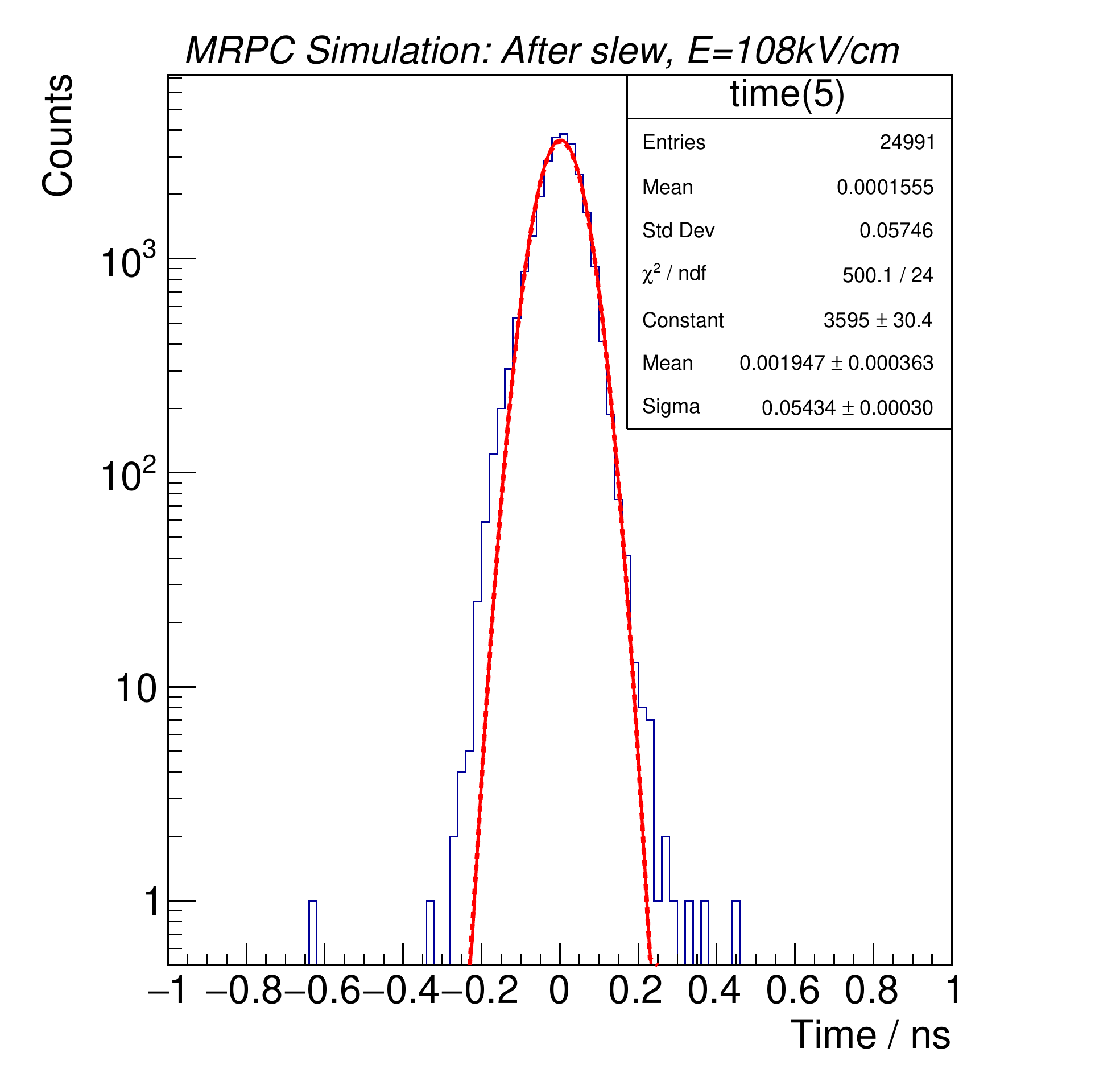}
        \caption{}
        \label{fig:slewc}
    \end{subfigure}
    \begin{subfigure}[b]{0.45\textwidth}
        \includegraphics[width=\textwidth]{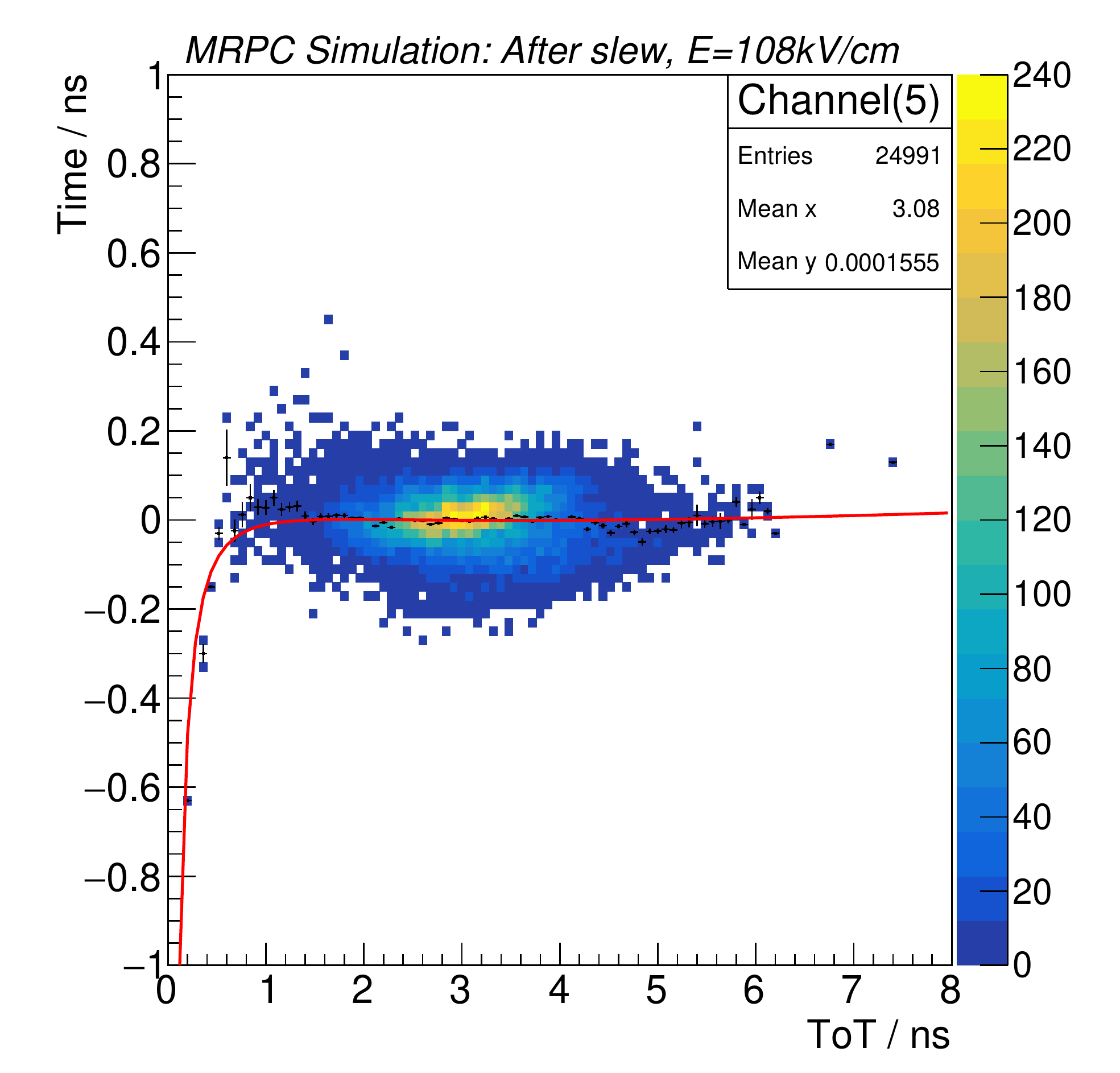}
        \caption{}
        \label{fig:slewd}
    \end{subfigure}
    \caption{The slewing correction. (a)(c) The distribution of $t_c$ before and after the slewing correction. (b)(d) 2D distribution of $t_{tot}$ versus $t_c$ before and after the correction,}
    \label{fig:slew}
\end{figure}

\section{Results of the simulation and comparison with experimental data}
\label{sec:ressim}
The detector performance at various electric fields and two kinds of the gas mixture is studied using the simulation presented here. A gaussian electronics noise of about $1/3$ of the threshold is added to the readout signal shown in Fig.\ref{fig:signal}(b). A fixed threshold of 20 fC is chosen and the standard deviation of the threshold crossing time $t_c$ is commonly defined as the MRPC time resolution. Due to the fixed threshold, signals with larger pulse height have earlier $t_c$ and vice versa. This time walk is only related with the signal amplitude and therefore should be corrected off-line with $t_{tot}$, like it was done in \cite{akindinov2004results}. This is called the slewing correction, and in this paper, the relationship of $t_c$ and $t_{tot}$ is fit and corrected with the function:
\begin{equation}
\label{eq:FEE}
t_c=a_0+\frac{a_1}{\sqrt{t_{tot}}}+\frac{a_2}{t_{tot}}+a_3t_{tot}+a_4t_{tot}^2+a_5t_{tot}^3
\end{equation}
Fig.\ref{fig:slew} shows the result of a simulated MRPC working with a gas mixture of 90\% $\rm C_2H_2F_4$, 5\% $\rm C_4H_{10}$ and 5\% $\rm SF_6$. The electric field in the gap is 108 kV/cm, and the efficiency at this condition is over 99\% and in the plateau region. Fig.\ref{fig:slew}(a) and (b) are the 1D distribution of $t_c$ and 2D distribution of $t_{tot}$ versus $t_c$ before the slewing correction, while Fig.\ref{fig:slew}(c) and (d) are the distributions after the correction. These plots prove that the dependence of $t_{tot}$ and $t_c$ is largely eliminated by the correction, and the final time resolution for this MRPC is 54 ps.   
\begin{figure*}[h!]
	\centering
	\includegraphics[width=0.4\textwidth]{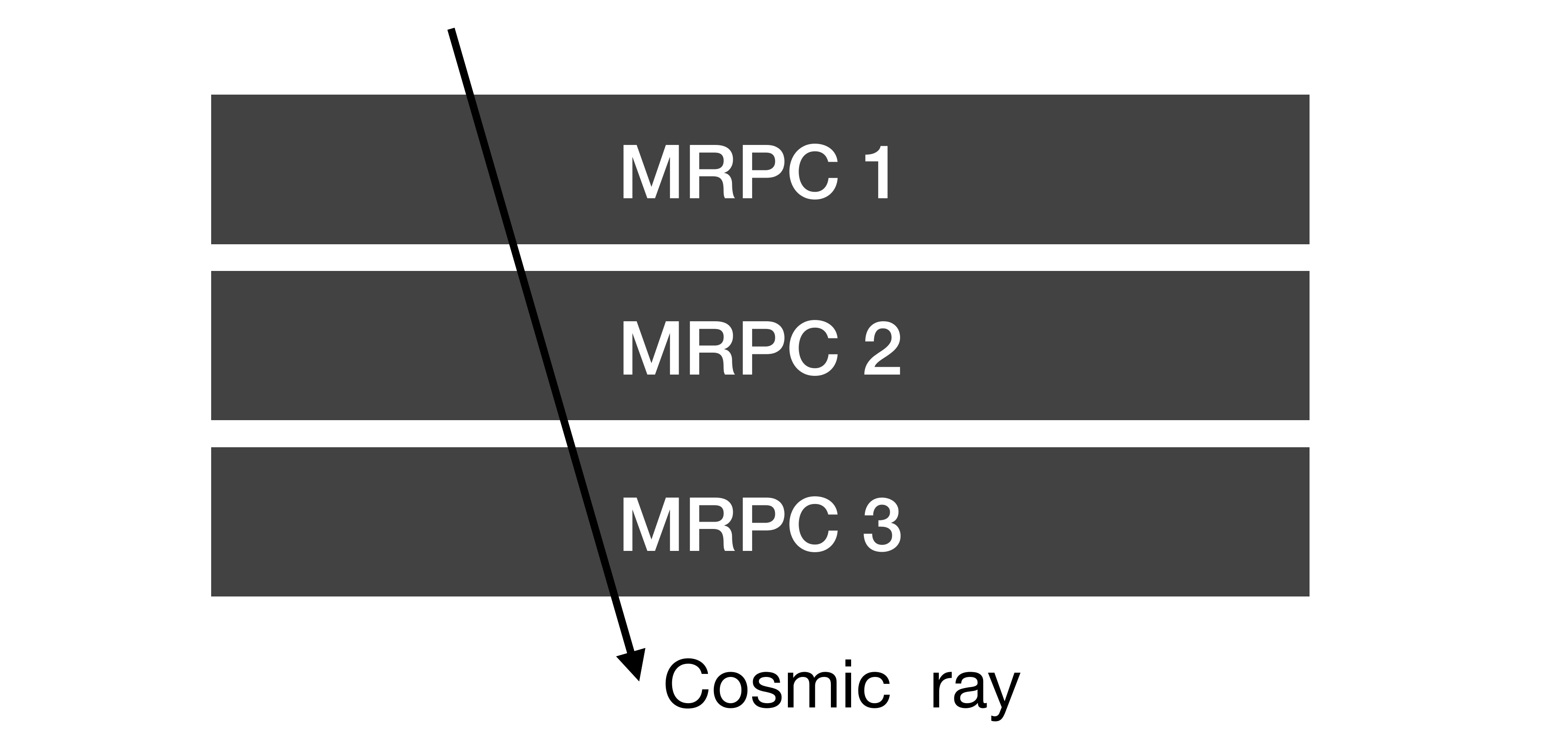}
	\caption{The setup of the experiment.}
	\label{fig:experset}
\end{figure*}
\begin{figure}
    \centering
    \begin{subfigure}[b]{0.45\textwidth}
        \includegraphics[width=\textwidth]{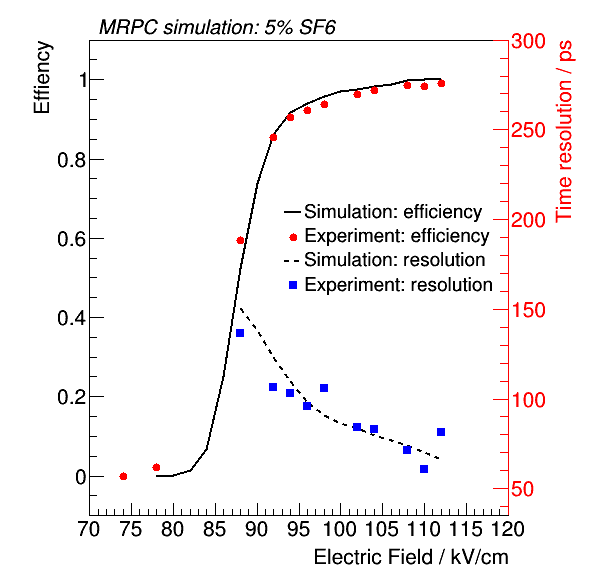}
        \caption{}
        \label{fig:scan1}
    \end{subfigure}
    \begin{subfigure}[b]{0.45\textwidth}
        \includegraphics[width=\textwidth]{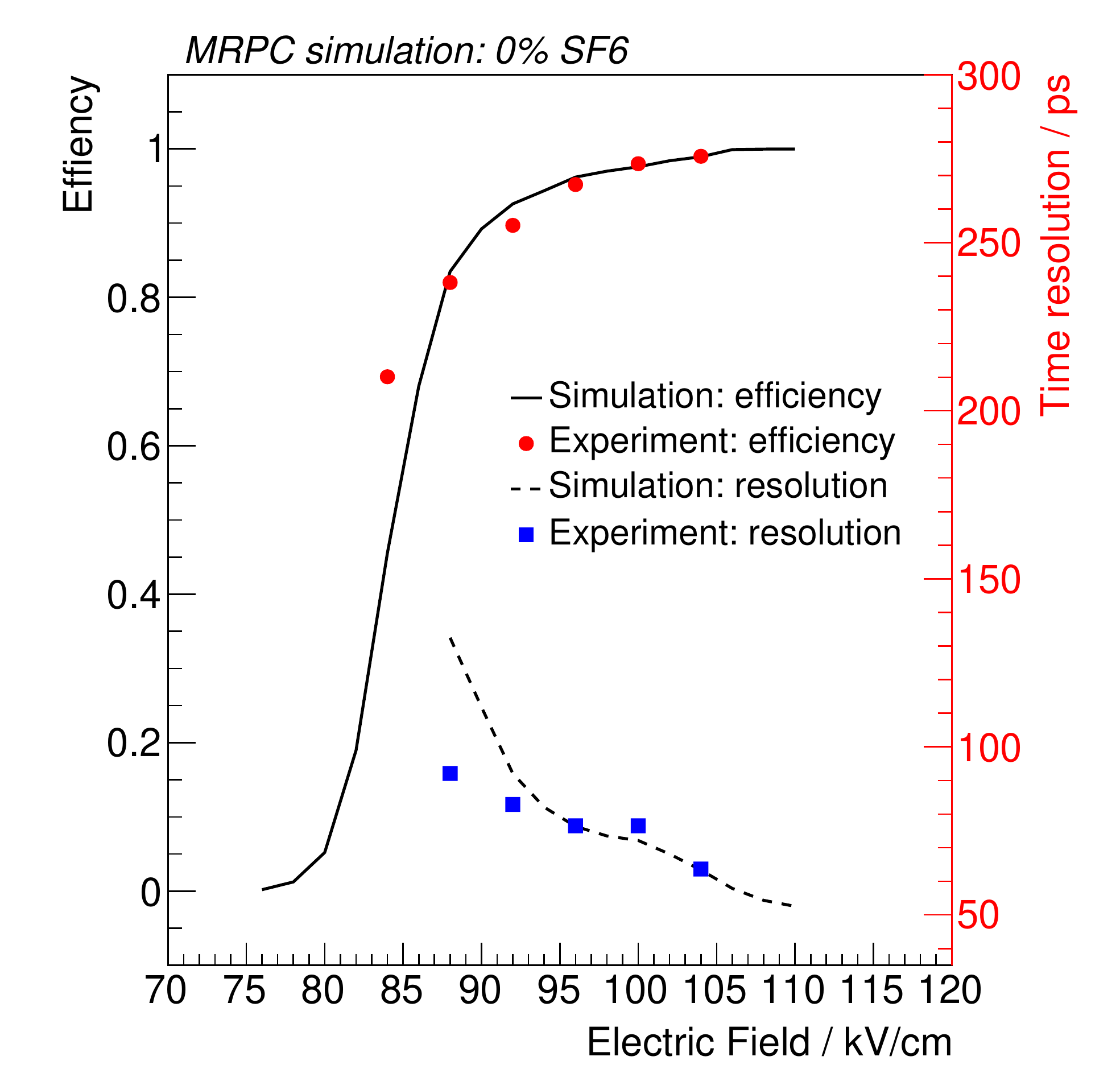}
        \caption{}
        \label{fig:scan2}
    \end{subfigure}
    \caption{The efficiency and time resolution of the MRPC with respect to the electric field with the gas mixture (a) 90\% $\rm C_2H_2F_4$, 5\% $\rm C_4H_{10}$ and 5\% $\rm SF_6$ and (b) 95\% $\rm C_2H_2F_4$ and 5\% $\rm C_4H_{10}$. The solid and dashed lines are the results from the simulation, while the circle and the square dots are from the experiment.}
    \label{fig:scan}
\end{figure}

To evaluate the effectiveness of the simulation, a cosmic ray test is conducted in the lab and the results are compared. Three identical MRPCs that have the same geometry as shown in Fig.\ref{fig:geo} are placed in an aluminum box and aligned vertically as shown in Fig.\ref{fig:experset}. The signals from the detectors are collected and amplified with the PADI electronics\cite{padi2014}, and the slewing correction for $t_c$ is the same as that for the simulation data. For the experiment setup, the efficiency is defined to be:
\begin{equation}
\label{eq:eff}
efficiency=\frac{N_{MRPC2}}{N_{MRPC1\&\&MRPC3}}
\end{equation}
\noindent $N_{MRPC2}$ is the number of events that are detected by MRPC2 and $N_{MRPC1\&\&MRPC3}$ is by both MRPC1 and MRPC3. The time difference of MRPC1 and MRPC2 is analyzed and since all the 3 MRPCs are identical, the MRPC time resolution is:
\begin{equation}
\label{eq:res}
time\ resolution=\frac{\sigma(t_2-t_1)}{\sqrt{2}}
\end{equation}
\noindent 
We scan the electric field from 74 kV/cm to 112 kV/cm for both the simulation and experiment. Fig.\ref{fig:scan} shows the result for efficiency and time resolution with respect to the electric field with two different gas mixtures (a) 90\% $C_2H_2F_4$, 5\% $C_4H_{10}$ and 5\% $SF_6$ and (b) 95\% $C_2H_2F_4$ and 5\% $C_4H_{10}$. The solid and dashed lines are the results from the simulation, while the circle and the square dots are from the experiment. This proved that the simulation agrees well with the experiment for both gas mixtures. The best time resolution we achieve with this MRPC is around 50$\sim$60 ps.

\section{Conclusions}
\label{sec:concl}
We have built a standalone simulation framework of the MRPC detectors and proven that the results agree well with the cosmic ray experiment. As the beam energy for the physics experiments is becoming larger and larger, it is necessary to improve both the detector structure and data analysis method. This framework can be used to optimize the detector design at different electric fields and with different gas mixtures. Meanwhile, the waveform about the signal is obtained from the simulation. It contains much more information of the signal than the previous $t_c$ and $t_{tot}$, and therefore new analysis method can be developed based on the waveform to improve time resolution.  

\section{Acknowledgments}
The work is supported by National Natural Science Foundation of China under Grant No.11420101004, 11461141011, 11275108, 11735009. This work is also supported by the Ministry of Science and Technology under Grant No. 2015CB856905, 2016 YFA0400100.
%This work was supported by the U.S.~Department of Energy, Office of Science under contract DE-AC02-05CH11231 and by the China Scholarship Council. 

%\bibliographystyle{ieeetr}
\bibliographystyle{elsarticle-num}
\bibliography{reference.bib}{}

\begin{thebibliography}{10}
\expandafter\ifx\csname url\endcsname\relax
  \def\url#1{\texttt{#1}}\fi
\expandafter\ifx\csname urlprefix\endcsname\relax\def\urlprefix{URL }\fi
\expandafter\ifx\csname href\endcsname\relax
  \def\href#1#2{#2} \def\path#1{#1}\fi

\bibitem{elvira2017impact}
V.~D. Elvira, Impact of detector simulation in particle physics collider
  experiments, Physics Reports 695 (2017) 1--54.

\bibitem{Atlasneural}
{ATLAS collaboration}, A neural network clustering algorithm for the atlas
  silicon pixel detector, JINST 9~(09) (2014) P09009.

\bibitem{egs2006}
R.~L. Ford, et~al., The egs code system: Computer programs for the monte carlo
  simulation of electromagnetic cascade showers (version 3), Tech. rep., SLAC
  National Accelerator Laboratory (SLAC) (2006).

\bibitem{geant1987}
R.~Brun, et~al., Geant 3: user's guide geant 3.10, geant 3.11, Tech. rep., CERN
  (1987).

\bibitem{zeballos1996avalanche}
E.~C. Zeballos, et~al., Avalanche fluctuations within the multigap resistive
  plate chamber, Nucl. Instrum. Meth. A381~(2-3) (1996) 569--572.

\bibitem{abbrescia1997properties}
M.~Abbrescia, et~al., Properties of c2h2f4-based gas mixture for avalanche mode
  operation of resistive plate chambers, Nucl. Instrum. Meth. A398~(2-3) (1997)
  173--179.

\bibitem{abbrescia1999progresses}
M.~Abbrescia, et~al., Progresses in the simulation of resistive plate chambers
  in avalanche mode, Nuclear Physics B-Proceedings Supplements 78~(1-3) (1999)
  459--464.

\bibitem{riegler500144}
W.~Riegler, et~al., Detector physics and simulation of resistive plate
  chambers, Nucl. Instrum. Meth. A500~(1-3) (2003) 144--162.

\bibitem{an2016monte}
F.~An, et~al., Monte-carlo study of the mrpc prototype for the upgrade of
  besiii, JINST 11~(09) (2016) C09022.

\bibitem{Agostinelli:2002hh}
{GEANT4 Collaboration}, {GEANT4: A Simulation toolkit}, Nucl. Instrum. Meth.
  A506 (2003) 250--303.

\bibitem{pdg4GeVmuon}
C.~Patrignani,
  \href{http://pdg.lbl.gov/2017/reviews/rpp2017-rev-cosmic-rays.pdf}{30.1.
  primary spectra}.
\newline\urlprefix\url{http://pdg.lbl.gov/2017/reviews/rpp2017-rev-cosmic-rays.pdf}

\bibitem{wang2010development}
J.~Wang, et~al., Development of multi-gap resistive plate chambers with
  low-resistive silicate glass electrodes for operation at high particle fluxes
  and large transported charges, Nucl. Instrum. Meth. A621~(1-3) (2010)
  151--156.

\bibitem{CBMTDR}
{CBM collaboration}, Technical design report for the cbm time-of-flight system.

\bibitem{ALLISON2016186}
J.~Allison, et~al., {Recent developments in Geant4}, Nucl. Instrum. Meth. A835
  (2016) 186--225.

\bibitem{Wang20181}
F.~Wang, et~al., The impact of incorporating shell-corrections to energy loss
  in silicon, Nucl. Instrum. Meth. A899 (2018) 1--5.

\bibitem{Apostolakis:2000yu}
J.~Apostolakis, et~al., An implementation of ionisation energy loss in very
  thin absorbers for the geant4 simulation package, Nucl. Instrum. Meth.
  A453~(3) (2000) 597--605.

\bibitem{Allison:1980vw}
W.~Allison, J.~Cobb, {Relativistic Charged Particle Identification by Energy
  Loss}, Ann. Rev. Nucl. Part. Sci. 30 (1980) 253--298.

\bibitem{townsend1900}
J.~Townsend, The conductivity produced in gases by the motion of
  negatively-charged ions, Nature 62~(1606) (1900) 340.

\bibitem{magboltz1997}
S.~Biagi, Magboltz, program to compute gas transport parameters, Version2.2,
  CERN.

\bibitem{lippmannthesis2003}
C.~Lippmann, Detector physics of resistive plate chambers, Ph.D. thesis,
  Frankfurt U. (2003).

\bibitem{spacecharge1998streamer}
P.~Camarri, et~al., Streamer suppression with sf6 in rpcs operated in avalanche
  mode, Nucl. Instrum. Meth. A414~(2-3) (1998) 317--324.

\bibitem{ramo1939}
S.~Ramo, Currents induced by electron motion, Proceedings of the IRE 27~(9)
  (1939) 584--585.

\bibitem{akindinov2004results}
A.~Akindinov, et~al., Results from a large sample of mrpc-strip prototypes for
  the alice tof detector, Nucl. Instrum. Meth. A532~(3) (2004) 611--621.

\bibitem{padi2014}
M.~Ciobanu, et~al., Padi, an ultrafast preamplifier-discriminator asic for
  time-of-flight measurements, IEEE transactions on nuclear science 61~(2)
  (2014) 1015--1023.

\end{thebibliography}

\end{document}